\documentclass[12pt,a4paper]{article}

\usepackage{graphicx}
\newcommand{\nn}{\nonumber}
\newcommand{\be}{\begin{equation}}
\newcommand{\ee}{\end{equation}}
\newcommand{\ba}{\begin{eqnarray}}
\newcommand{\ea}{\end{eqnarray}}
\newcommand{\ci}[1]{\cite{#1}}
\def\gev{\,{\rm GeV}}
\newcommand{\tw}{\textwidth}
\newcommand{\req}[1]{(\ref{#1})}
\def\={\,=\,}
\newcommand{\lsim}{\raisebox{-4pt}{$\,\stackrel{\textstyle <}{\sim}\,$}}

\begin{document}
\textwidth=135mm
 \textheight=200mm
\begin{center}
{\bfseries Generalized parton distributions from meson leptoproduction~\footnote
{{\small Talk presented at SPIN12, JINR, Dubna, September 17 - 22,
2012.}}}
\vskip 5mm
P.~Kroll$^{\dag}$ 
\vskip 5mm
{\small {\it $^\dag$ 
Fachbereich Physik, Universit\"at Wuppertal, D-42097 Wuppertal,
Germany}}
\end{center}
\vskip 5mm
\centerline{\bf Abstract}
Generalized parton distributions (GPDs) extracted from exclusive meson leptoproduction
within the handbag approach are briefly reviewed. Only the GPD $E$ is discussed 
in some detail. Applications of these GPDs to virtual Compton scattering (DVCS) and to
Ji's sum rule are also presented.  
\vskip 10mm
\section{Introduction}
The handbag approach to hard exclusive leptoproduction of photons and 
mesons (DVMP) off protons has extensively been studied during the last fifteen
years. The handbag approach is based on factorization of the process amplitudes 
in a hard subprocess, e.g.\ $\gamma^*q\to \gamma (M) q$, and soft hadronic matrix 
elements parametrized in terms of GPDs. This factorization property has been 
shown to hold rigorously  in the generalized Bjorken regime of large photon 
virtuality, $Q$, and large energy $W$ but fixed $x_B$. Since most data, in 
particular the data from Jlab, are not measured in this kinematical regime one 
has to be aware of power corrections from various sources. Which kind of power 
corrections are the most important ones and have to be taken into account is still
under debate. Nevertheless progress has been made in the understanding of the DVCS 
and DVMP data. In this talk I am going to report on an extraction of the GPDs from 
DVMP \ci{GK1}. In this analysis the GPDs are constructed from double 
distributions (DDs)\ci{mueller94, rad98} and the subprocess amplitudes are 
calculated taking into account quark transverse degrees of freedom as well as Sudakov 
suppression \ci{li}. The emission and reabsorption of the partons by the 
protons are treated collinearly. This approach also allows to calculate the
amplitudes for transversely polarized photons 
which are infrared singular in collinear factorization. The transverse photon amplitudes
are rather strong for $Q^2\lsim 10\,\gev^2$ as is known from the ratio of longitudinal
and transverse cross sections \ci{h1}. I am also going to discuss 
several applications of the extracted set of GPDs like the calculation of DVCS
observables from them \ci{kms} or the evaluation of the parton angular momenta. 
\section{The double distribution representation}
There is an integral representation of the GPD
$F^i = H^i, E^i, \widetilde{H}^i, \ldots$ ($i=u, d, s, g$)  in terms of DDs 
\ci{mueller94,rad98}
\be
F^i(x,\xi,t)\=\int_{-1}^{1}\, d\rho\,\int_{-1+|\rho|}^{1-|\rho|}\, d\eta
              \,\delta(\rho+\xi\eta -x)\, f_i(\rho,\eta,t)
              + D_i(x,t)\, \Theta(\xi^2-x^2)\,,
\label{eq:int-rep}
\ee  
where $D$ is the so-called $D$-term
\ci{pol99} which appears for the gluon and flavor-singlet quark combination of
the GPDs $H$ and $E$. The advantage of the DD representation is that polynomiality 
of the GPDs is automatically satisfied. A popular ansatz for the DD is 
\be
f_i(\rho,\eta,t) \= F^i(\rho,\xi=0,t)\, w_i(\rho,\eta)\,.
\label{eq:DD}
\ee
where the weight function $w_i$ that generates the skewness dependence of the
GPD, is assumed to be 
\be
w_i(\rho,\eta)\= \, \frac{\Gamma (2n_i+2)}{2^{2n_i+1}\Gamma^2 (n_i+1)} \,
\frac{[(1-|\rho |)^2-\eta^2]^{n_i}}{(1-|\rho |)^{2n_i+1}} 
\label{eq:weight}
\ee   
(in \ci{GK1} $n=1$ for valence quarks and 2 for sea quarks and gluons).
The zero-skewness GPD is parametrized as its forward limit multiplied
by an exponential in Mandelstam $t$
\be
F^i(\rho,\xi=0,t) \= F^i(\rho,\xi=0,t=0)\, \exp \big[ t p_{fi}(\rho) \big]
\label{eq:zero-skewness}
\ee
The profile function, $p_{fi}(\rho)$, is parametrized in a Regge-like manner
\be
p_{fi}(\rho) \= -\alpha_{fi}^\prime \ln{\rho} + B_{fi}
\label{eq:profile}
\ee
where $\alpha^\prime$ represents the slope of an appropriate Regge trajectory
and $B$ parametrizes the $t$ dependence of its residue. This profile 
function is a simplified version of a more complicated one that has been proposed in
\ci{DFJK4} 
\be
p_{fi}(\rho) \= \big(\alpha_{fi}^\prime \ln{1/\rho} + B_{fi}\big)\,(1-\rho)^3 
             + A_{fi}\,\rho(1-\rho)^2
\label{eq:profile-dfjk4}
\ee
The profile function \req{eq:profile} approximates \req{eq:profile-dfjk4} for
small $\rho$. Because of a strong $\rho - t$ correlation observed in \ci{DFJK4}
the profile function \req{eq:profile} can be applied at small $-t$. For the forward 
limits of the GPDs $H$, $\widetilde{H}$ and $H_T$ the corresponding parton 
distributions (PDFs) are used. The forward limits of the other GPDs are parametrized 
in a fashion analogously to the PDFs  
\be
  F^i(\rho,\xi=t=0) \= c_i \rho^{-\alpha_i} (1-\rho)^{\beta_i}
\label{eq:forward}
\ee
For alternative parametrizations of the GPDs in terms of conformal SL(2,R) partial 
waves see \ci{MM,kmls}.
\section{Extraction of the GPDs}
The GPDs are inserted into the convolution formula for vector mesons
\be
{\cal F}^{\,i}_{\rm V}(\xi, t, Q^2) \= \sum_\lambda \int^1_{x_i} dx 
            {\cal A}^i_{0\lambda, 0\lambda}(x,\xi,Q^2,t=0)\, F^i(x,\xi,t)
\label{eq:convolutions-mesons}
\ee
where $i=g, q$, $x_g=0$, $x_q=-1$ and $F$ either $H$ or $E$. A similar
convolution formula holds for pseudoscalar mesons. The subprocess amplitude ${\cal A}$
for partonic helicity $\lambda$ is to be calculated perturbatively using
$k_\perp$ factorization. In collinear factorization the cross section
for the production of vector mesons drops as $1/Q^6(\log{Q^2})^n$  with increasing 
$Q^2$ while experimentally \ci{h1} it approximately falls as $1/Q^4$. 
The required suppression of the amplitudes at low $Q^2$ is generated by the evolution
of the GPDs and by $k_\perp$ effects.
In \ci{MM} however GPDs are proposed which have a much stronger evolution.
At least for HERA kinematics these GPDs lead to fair fits of the HERA data on vector
meson electroproduction in collinear factorization. 

In \ci{GK1} parameters of the double distribution are fitted to the available
data on $\rho^0$, $\phi$ and $\pi^+$ production from HERMES, COMPASS, E665, H1
and ZEUS. The data cover a large kinematical range: $3\,\gev^2\lsim Q^2\lsim 100\,\gev^2$, 
$4\,\gev\lsim W\lsim 180\,\gev$. Data from the present Jlab are not taken into 
account in these fits because they are likely affected by strong power corrections 
at least in some cases (e.g. $\rho^0$ production). Constraints from nucleon form
factors and from positivity bounds \ci{DFJK4} are taken into account.
Some parameters of the transversity GPDs needed for pion electroproduction, are fixed 
by lattice QCD results \ci{lattice}. The fit leads to a fair
description of all the mentioned data. What we learned about the GPDs is
summarized in Tab.\ \ref{tab1}.
 
\begin{table}[t]
\renewcommand{\arraystretch}{1.2} 
\begin{center}
\begin{tabular}{| c || c | c | c |}
\hline   
GPD &  probed by &  constraints &  status \\[0.2em]
\hline
$H$(val) & {\small $\rho^0, \phi$ cross sections} & {\small PDFs, Dirac ff} & {\small ***} \\[0.2em]
$H$(g,sea) & {\small $\rho^0, \phi$ cross sections} & {\small PDFs} & {\small ***} \\[0.2em]
$E$(val)& {\small $A_{UT}(\rho^0, \phi)$} & {\small Pauli ff} & {\small **} \\[0.2em]
$E$(g,sea) &  - & {\small sum rule for $2^{nd}$ moments}
& {\small {-}}\\[0.2em]
$\widetilde{H}$ (val)  & {\small $\pi^+$ data} &{\small pol.\ PDFs, axial ff} &
{\small **} \\[0.2em]
$\widetilde{H}$(g,sea)  & $A_{LL}(\rho^0)$  & {\small polarized PDFs} & {\small
    *}\\[0.2em]
$\widetilde{E}$ (val)& {\small $\pi^+$ data} & pseudoscalar ff  
                                          & {\small *} \\[0.2em]
$H_T, \bar{E}_T$(val) & {\small $\pi^+$ data} & {\small transversity PDFs} 
                                          & {\small *}\\[0.2em]
\hline
\end{tabular}
\caption{\label{tab1} Status of small-skewness GPDs as extracted from meson
leptoproduction data. At present no information is available on GPDs not 
appearing in the list. Except of $H$ for gluons and sea quarks all GPDs are
only probed for scales of about $4\,{\rm GeV}^2$. For comparison five stars 
are assigned to PDFs.} 
\end{center}
\renewcommand{\arraystretch}{1.0}   
\end{table}

\section{DVCS}
Because of universality the GPDs extracted from DVMP, may for instance be
applied to DVCS. This process is 
calculated in \ci{kms} to leading-twist accuracy and leading-order of pQCD 
while the Bethe-Heitler (BH) contribution is worked out without any approximation. 
The DVCS part involves convolution integrals, so-called Compton form factors, which read
\be
{\mathcal F}(\xi,t)\= \int_{-1}^1 dx \Big[e_u^2 F^u+ e_d^2 F^d + e_s^2 F^s\Big]
                       \,\Big[\frac1{\xi-x-i\varepsilon}-\epsilon_f\,
                       \frac1{\xi+x-i\varepsilon}\Big]\,,
\label{eq:CFF-def}
\ee
where $\epsilon_f=+1$ for $F=H, E$ and $-1$ for $\widetilde{H}, \widetilde{E}$.  
With the GPDs at disposal the convolution integrals can evaluated free of
parameters. The corresponding amplitudes for DVCS can be combined with the 
BH amplitudes in those for leptoproduction of photons
\begin{equation}
| \mathcal{M}_{l p \rightarrow l p \gamma} |^2 \= | \mathcal{M}_{\rm{BH}}|^2 
      + \mathcal{M}_{\rm{I}} + | \mathcal{M}_{\rm{DVCS}} |^2\,.
\label{eq-bh-interf-vcs}
\end{equation}
The three terms in \req{eq-bh-interf-vcs} have the following harmonic
structure in $\phi$, the azimuthal angle of the outgoing photon with regard to
the leptonic plane (i=BH, DVCS, interference):
\be
|\mathcal{M}_i|^2  \propto  L_i\sum_{n=0}^3 \left[ c_n^i \cos (n\phi ) + 
      s_n^i \sin (n\phi ) \right] \, ,
\label{eq-cross-section} 
\ee
\noindent where $L_i=1/[-t P(\cos\phi)]$ for the Bethe-Heitler and the
interference term and $L_{\rm DVCS}=1$. Although there are only harmonics up
to the maximal order 3 in the sums, the additional $\cos\phi$ dependence from
the lepton propagators, included in $P(\cos{\phi})$, generates in principle an 
infinite series of harmonics for the BH and interference terms. A more
detailed harmonic structure taking into account beam and target polarizations
can be found for instance in \ci{diehl-sapeta}.

\begin{figure}[ht]
\begin{center}
\includegraphics[width=0.33\tw]{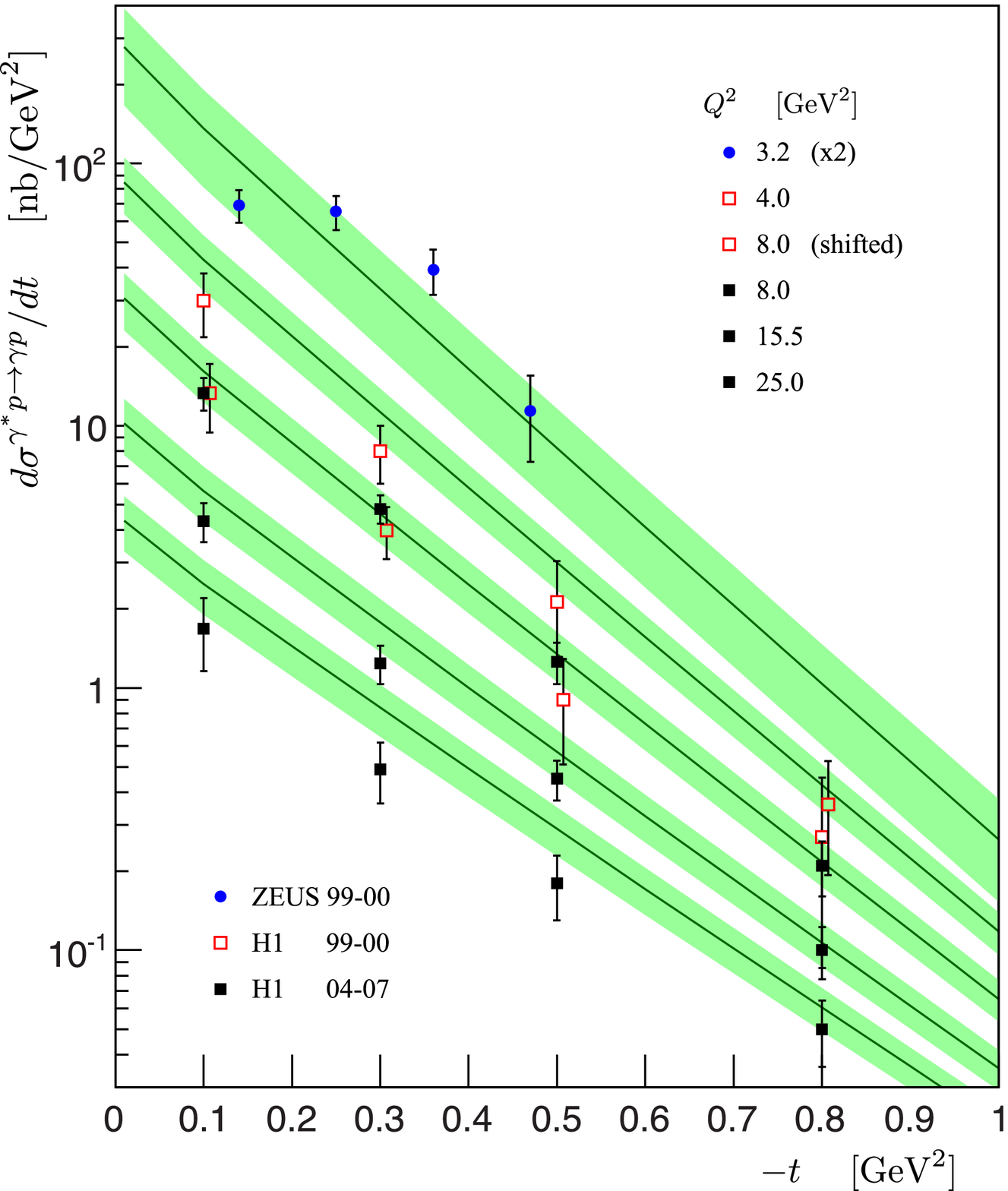}\hspace*{0.08\tw}
\includegraphics[width=.50\tw]{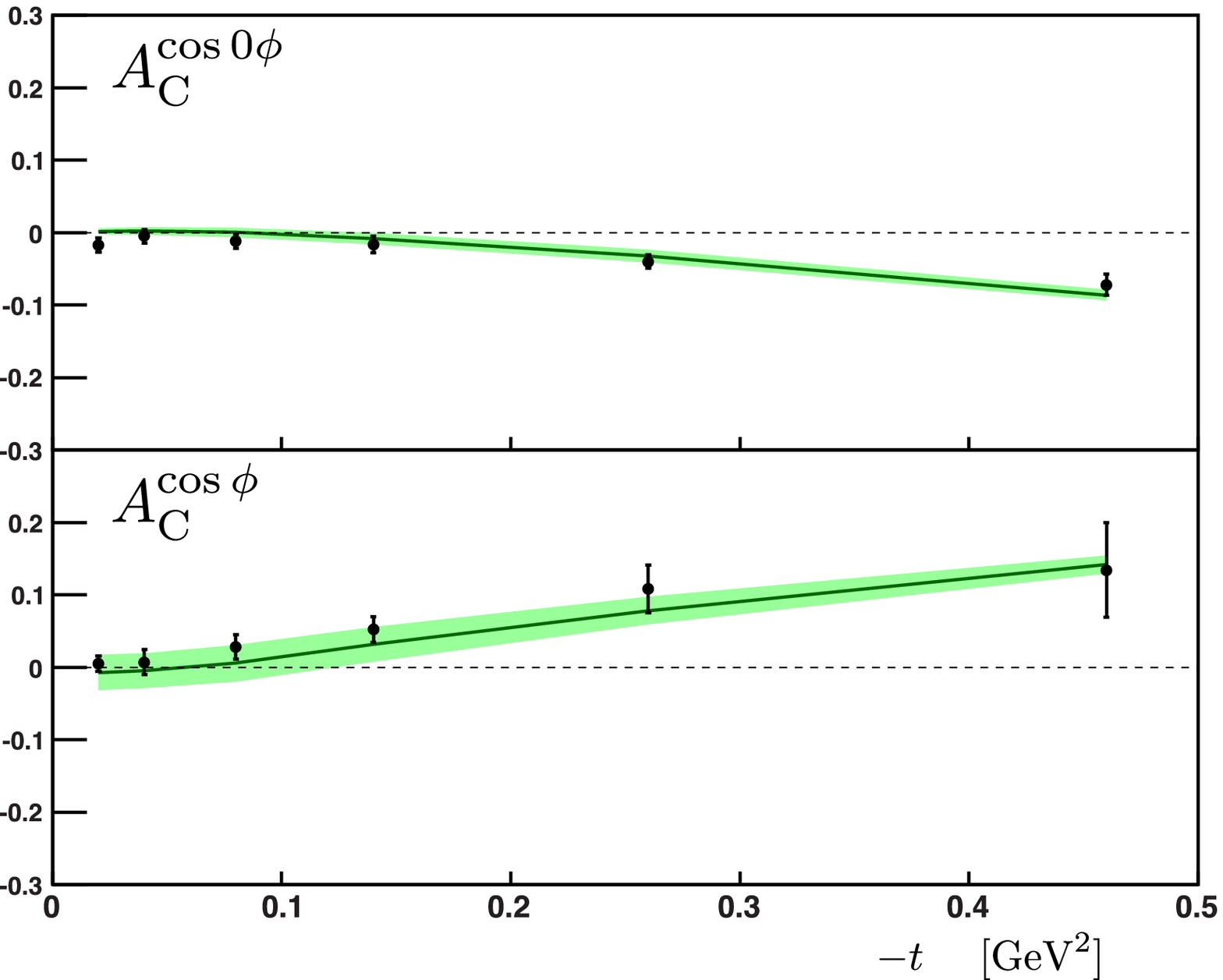}
\caption{The DVCS cross section (left) and the beam charge asymmtry (at 
$Q^2\simeq 2.51\,\gev^2$, $x_B\simeq 0.097$. Data are taken from \ci{H1,zeus} and 
\ci{hermes-bca}. The theoretical results obtained in \ci{kms} are shown as solid
lines with shadowed bands representing their uncertainties.}   
\label{fig:dvcs}
\end{center}
\end{figure}

A detailed comparison of this theoretical approach with experiments performed
in \ci{kms}, reveals reasonable agreement with HERMES, H1 and ZEUS data and
a less satisfactory description of the large-skewness, small $W$ Jlab data. 
The GPDs extracted in \ci{GK1} are not optimized for this kinematical region. 
As examples the DVCS cross section at HERA
kinematics and the beam charge asymmetry are shown in Fig.\ \ref{fig:dvcs}.
It should be mentioned that, in the same spirit, a DVCS analysis is performed
in \ci{MM,kmls} 

\section{What do we know about the GPD $E$?}
Let me now discuss the GPD $E$ in some detail. The analysis of the nucleon form factors
performed in \ci{DFJK4} provided the zero-skewness GPDs for valence quarks with the
profile function \req{eq:profile-dfjk4} which can be used as input to the 
DD representation \req{eq:int-rep}. Since in 2004 data on the neutron
form factors were only available for $Q^2\lsim 2\,\gev^2$ the parameters of the zero-
skewness GPD $E$ were not well fixed; a wide range of values were allowed
for the powers $\beta^u_e$ and $\beta^d_e$. There is an ongoing reanalysis of the
form factors \ci{dk12} making use of the new data which for the neutron now 
extend to much larger values of $Q^2$. The new results for the valence-quark GPDs 
are similar to the 2004 ones but the powers $\beta^q_e$ are better determined.

Not much is known about $E^g$ and $E^{\rm sea}$. 
There is only a sum rule for the second moments of $E$ \ci{diehl-kugler} at 
$t=\xi=0$
\be
\int_0^1 dx E^g(x,\xi=0,t=0) \= e_{20}^g\=-\sum e_{20}^{q_v} -2\sum e_{20}^{\bar{q}}\,.
\label{eq:SumRuleE}
\ee 
It turns out that the valence contribution to the sum rule is very small. Hence,
the second moments of the gluon and sea-quark GPD $E$ cancel each other almost completely.
Since the parametrization \req{eq:forward} for the forward limit of $E$ 
does not have nodes except at the end-points this property approximately holds of other 
moments as well and even for the convolution \req{eq:convolutions-mesons}.

For $E^s$ a positivity  bound for its Fourier transform with respect to the momentum 
transfer \ci{DFJK4} forbids a large strange quark contribution and, assuming a 
flavor-symmetric sea, a large gluon contribution too:
\be
\frac{b^2}{m^2}\left(\frac{\partial e_s(x,b)}{\partial b^2}\right) \leq s^2(x,b)-\Delta s^2(x,b)
\ee
where $s$, $\Delta s$ and $e_s$ are Fourier conjugated to $H^s$, $\widetilde{H}^s$ and $E^s$,
respectively. The impact parameter ${\bf b}$ is canonically conjugated to the 
two-dimensional momentum transfer ${\bf \Delta}$ ($\Delta^2=t$). The bound on $E^s$ is 
saturated for $c_e^s=\pm 0,155$ ($\beta_e^s=7$ in \req{eq:forward}) \ci{GK4}.
The normalization of $E^g$ can subsequently be fixed from the sum rule \req{eq:SumRuleE} 
($\beta^g_e=6$). These results are inserted in \req{eq:int-rep} in order to obtain an 
estimate of $E^{\rm sea}$ and $E^g$.

\begin{figure}[t]
\begin{center}
\includegraphics[width=0.7\tw]{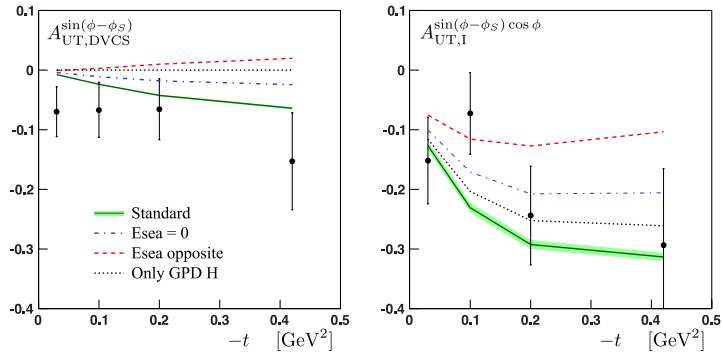}
\caption{The transverse target asymmetries for DVCS and the BH-DVCS interference.
Data are taken from \ci{hermes-aut-dvcs}, theoretical results from \ci{kms}.} 
\label{fig:AUT}
\end{center}
\end{figure}

The GPD $E$ is probed by transverse target asymmetries 
\be 
 A_{UT} \sim {\rm Im}\Big[{\mathcal E}^* {\mathcal H}\Big]\,,
\ee
for given $H$ as extracted from the cross sections of vector-meson leptoproduction 
\ci{GK1}. The data on $\rho^0$ production from HERMES \ci{hermes-aut}
and COMPASS \ci{compass-aut} are well fitted by the described parametrization of $E$.
However, only $E$ for valence quarks matters
for $A_{UT}$ since the sea and gluon contribution to $E$ cancel to a large extent.
Fortunately the analysis of DVCS data \ci{kms} provides additional although not very
precise information on $E^{\rm sea}$. To leading-oder of pQCD there is no gluon contribution
in DVCS and therefore $E^{\rm sea}$ becomes visible. The HERMES collaboration
has measured the transverse target asymmetries for DVCS and for the BH-DVCS interference 
term \ci{hermes-aut-dvcs}. The data are shown in Fig.\ \ref{fig:AUT} and compared to
the results obtained in \ci{kms}. Despite the large experimental errors it
seems that a negative $E^{\rm sea}$ is favored. 
Independent information on $E^g$ would be of interest. This may be 
obtained from a measurement of the transverse target polarisation in $J/\Psi$
photoproduction \ci{koempel} . 
       
\section{Summary}
I have briefly summarized the recent progress in the analysis for hard exclusive
leptoproduction of mesons and photons within the handbag approach. We learned
that the data on both reactions are consistent with each other in so far as they
can be described with a common set of GPDs. In fact the GPDs contructed from
double distributions and with parameters adjusted to the meson data allow for
a parameter-free calculation of DVCS.

The knowledge of the GPDs allow for an evaluation of the angular momenta the
partons inside the proton carry. At $\xi=t=0$ they are given by the second moments
of $H$ and $E$
\be
2J^a\=\Big[q^q_{20}+e_{20}^q\Big]\,, \qquad 2J^g\=\Big[g_{20}+e_{20}^g\Big]
\ee
Taking the values of the $H$-moments are taken from the CTEQ6 PDFs \ci{cteq6}, those 
for $E$ from the form factor analysis \ci{DFJK4} and from the analysis of $A_{UT}$
for DVMP \ci{GK4} and DVCS \ci{kms}, one obtains at the scale $4\,\gev^2$ 
\ba
J^{u}&=& 0.250 \quad J^{d}= 0.020 \quad J^{s}=0.015 \quad
        J^g= 0.214 \quad (\mathrm{ for}\; E^s=E^g=0) \nn\\
 & & 0.225 \hspace*{0.065\tw} -0.005 \hspace*{0.0635\tw} -0.011 \hspace*{0.105\tw}
             0.286  \quad (\mathrm{ for}\; E^s<0, E^g>0)   \nn
\ea
The main uncertainty comes from the badly known $E^s$ contribution although it 
is strongly reduced due to the DVCS analysis which favors a negative $E^s$
while from DVMP alone it could also be positive. The large value of $J^g$ is no
surprise. The value of $g_{20}$ represents the familiar result that about
$40\%$ of the proton's momentum is carried by the gluons. Since $E^g$ seems to be positive
according to \ci{kms} it even increases $J^g$. New PDF analyses and the reanalysis
of the nucleon form factors \ci{dk12} will improve the results on $J$.

\end{document}